\documentclass[aps,twocolumn,nofootinbib, superscriptaddress, floatfix]{revtex4-1}

\usepackage{multirow}
\usepackage{booktabs}
\usepackage{placeins}
\usepackage{soul}
\usepackage{bbm}
\usepackage{amsmath,amssymb,amsbsy,bm,amsfonts,mathrsfs}
\usepackage{graphicx}
\usepackage{wrapfig}
\usepackage{url}
\usepackage{float}
\usepackage{lipsum}
\usepackage{enumitem}
\usepackage{titlesec}
\usepackage{microtype}

\usepackage[usenames,dvipsnames,svgnames]{xcolor}
\definecolor{redd}{rgb}{0.8, 0.1,0.2}
\definecolor{navy}{rgb}{0.05, 0.23,0.75}

\usepackage[colorlinks]{hyperref}
\hypersetup{
    colorlinks    =    true,
    citecolor     =    navy,
	linkcolor     =    redd,
	urlcolor      =    navy,
	anchorcolor   =    blue
}

\begin{document}

\title{Softly shifting away from dark matter direct detection}

\author{Chuan-Yang Xing}
\email{cyxing@pku.edu.cn}
\affiliation{Department of Physics and State Key Laboratory of Nuclear Physics and Technology, Peking University, Beijing 100871, China}

\author{Ling-Xiao Xu}
\email{lingxiaoxu@pku.edu.cn}
\email{lingxiao.xu@sns.it}
\affiliation{Department of Physics and State Key Laboratory of Nuclear Physics and Technology, Peking University, Beijing 100871, China}
\affiliation{Scuola Normale Superiore, Piazza dei Cavalieri 7, 56126, Pisa, Italy}

\author{Shou-hua Zhu}
\email{shzhu@pku.edu.cn}
\affiliation{Department of Physics and State Key Laboratory of Nuclear Physics and Technology, Peking University, Beijing 100871, China}
\affiliation{Collaborative Innovation Center of Quantum Matter, Beijing, 100871, China}
\affiliation{Center for High Energy Physics, Peking University, Beijing 100871, China}

\begin{abstract}

We propose soft breaking mechanism for dark matter (DM) shift symmetry in a class of composite dark matter models, where both DM and the Higgs boson arise as pseudo Nambu-Goldstone bosons from novel strong dynamics. Our mechanism is utilized to suppress the non-derivative portal coupling between the Higgs boson and DM particle, which can evade the stringent bound of current DM direct detection experiments. Otherwise this non-derivative portal coupling would naturally be at the same order of the Higgs quartic, rendering this class of models under severe crisis. For realizing soft breaking mechanism, we introduce vector-like top partners, dubbed as ``softons", to restore the shift symmetry of DM in top Yukawa sector, which however is only broken by the softon masses. The portal coupling would automatically vanish as the shift-symmetry-breaking softon masses approach zero. Specifically we present a proof-of-concept model of soft breaking, based on the coset $O(6)/O(5)$ and the simplest fermion embedding, and study its DM phenomenology, where we show a large amount of novel parameter space is opened up by using the soft breaking mechanism.

\end{abstract}

\maketitle

\section{Introduction} \label{sec:intro}

One plausible resolution to the hierarchy problem is to consider the Higgs as a pseudo Nambu-Goldstone boson (PNGB) emerging from spontaneous global symmetry breaking of a new strongly-interacting sector. In this scenario, Higgs is protected by an approximate shift symmetry and thus naturally lighter than the typical scale of the strong sector. The minimal realization~\cite{Agashe:2004rs,Contino:2006qr,Contino:2010rs} contains four Goldstone bosons forming the Higgs doublet. Electroweak symmetry breaking (EWSB) is triggered due to radiative corrections and the masses of the Higgs and EW gauge bosons are at the same scale.
The mass scale of the strong sector resonances can be uplifted beyond current collider limits~\cite{Blasi:2019jqc} and the scenario is still phenomenologically intriguing.
Besides, extra Goldstones arise if non-minimal symmetry breaking patterns are employed~\cite{Gripaios:2009pe,Mrazek:2011iu,DaRold:2019ccj}. Some of these Goldstones are natural dark matter (DM) candidates if they are stabilized by a discrete parity or dark $U(1)$ symmetry~\cite{Frigerio:2012uc,Chala:2012af,Marzocca:2014msa,Fonseca:2015gva,Balkin:2017aep,Ma:2017vzm,Balkin:2018tma,Davoli:2019tpx,Cacciapaglia:2019ixa,Ramos:2019qqa,Ahmed:2020hiw}, where the Higgs and DM are unified in the same framework with comparable masses and DM would exhibit typical phenomenology of weakly interacting massive particles (WIMP); see e.g.~\cite{Bertone:2004pz, Lisanti:2016jxe, Plehn:2017fdg, Profumo:2019ujg} for reviews.

Within this class of models, when DM candidates are gauge singlets of the Standard Model (SM), they interact with SM particles mainly through the Higgs portal. The leading operators characterizing the portal interactions between the Higgs doublet $H$ and the real singlet scalar DM $\eta$ are~\cite{Balkin:2018tma}
\begin{equation}
\mathcal{O}_1 = \frac{1}{f^2} \partial_\mu (H^\dagger H) \partial^\mu (\eta^2),
\quad
\mathcal{O}_2 = \lambda H^\dagger H \eta^2,
\label{operators}
\end{equation}
where $f$ denotes the symmetry breaking scale and $\lambda$ is the coupling strength of the usual Higgs portal interaction. The operator $\mathcal{O}_1$ characterizes the derivative portal coupling originating from the Goldstone nature of the Higgs and DM, while the operator $\mathcal{O}_2$ only arises from the radiative scalar potential of the Goldstones.
Phenomenologically, $\mathcal{O}_1$ is enhanced in the high energy regime and dominates over $\mathcal{O}_2$, e.g. in DM annihilation, although $\mathcal{O}_1$ has higher mass dimension. Nevertheless, $\mathcal{O}_2$ would dominate in direct detection experiments, where $\mathcal{O}_1$ is highly suppressed due to small momentum transfer~\cite{Barducci:2016fue}. Therefore, direct detection experiments already impose stringent bound on the coupling strength $\lambda$.
Following naive dimensional analysis (NDA), $\lambda$ is estimated to be of the same order of the Higgs quartic as $\frac{1}{2}\lambda_h \simeq 0.065$, if the top sector breaks both the shift symmetry of the Higgs and DM. However, this value of $\lambda$ is in tension with current direct detection bounds~\cite{Akerib:2016vxi,Cui:2017nnn,Aprile:2018dbl}. Motivated by this fact, top sector is required to fully respect the DM shift symmetry, which usually renders the top sector non-minimal.
Two different scenarios are proposed and discussed in Ref.~\cite{Balkin:2018tma} to evade direct detection bounds and $\lambda$ can be automatically suppressed if the leading shift-symmetry breaking effects arise in the bottom sector rather than the top sector, or a novel dark photon sector with all the quarks preserving the DM shift symmetry.

In this work, we propose a mechanism for softly breaking the DM shift symmetry in the top sector, leading to suppressed portal coupling $\lambda$ while DM still lives in the weak scale. In conventional top sectors, DM shift symmetry is broken since top quark does not fulfill complete representations of the global symmetry. In contrast, to realize soft-breaking, two model ingredients are as follows.
\begin{enumerate}
\item Extra vector-like top partners, dubbed as ``softons", are introduced to restore exact DM shift symmetry in top Yukawa terms.
\item Softon mass terms are naturally introduced, and they {\it{softly}} break DM shift symmetry, leading to nonzero DM mass and the portal coupling $\lambda$.
\end{enumerate}
The numerical value of the portal coupling $\lambda$ is proportional to the shift-symmetry-breaking softon masses.
At the limit of zero softon masses, DM particle remains an exact Goldstone boson and all the non-derivative terms of DM vanish. For TeV-scale softon masses, $\lambda$ can be small enough to evade the bounds of direct detection experiments.
In parallel, we note that similar idea is implemented to eliminate quadratic divergence in neutral-naturalness models~\cite{Xu:2018ofw} and stabilize the Higgs mass in composite Higgs models~\cite{Blasi:2019jqc,Blasi:2020ktl}. Also, in Ref.~\cite{Ahmed:2020hiw}, direct detection signal is reduced by giving vector-like masses to the symmetry partners of the top quark, in which these top partners are part of a neutral-naturalness construction. Here we will discusse the soft breaking mechanism in a more general framework, and demonstrate the idea in the top sector of a more minimal model.

\section{Soft Breaking Mechanism} \label{sec:soft-breaking}

As a proof-of-concept example, we focus on the next-to-minimal coset $SO(6)/SO(5)$~\cite{Gripaios:2009pe} to illustrate the soft-breaking mechanism for PNGB DM, and similar implementation can straightforwardly be generalized to other cosets. In the following, we explain in details the implementation in the top sector.

Within the unbroken global $SO(5)$ symmetry, the custodial symmetry is identified as $SU(2)_L \times SU(2)_R \cong SO(4)$ where $SU(2)_L \times U(1)_Y$ is further weakly gauged. In the unitary gauge, the PNGBs are explicitly
\begin{equation}
\Sigma = \frac{1}{f} ( 0, 0, 0, h, \eta, \sqrt{f^2-h^2-\eta^2} )^T
\label{eq:sig}
\end{equation}
transforming as a fundamental representation of $SO(6)$ global symmetry and satisfying the constraint $\Sigma^T\Sigma=1$. Within the field $\Sigma$, $\eta$ is a DM candidate if it is stabilized by symmetries, and this is achieved by imposing a dark parity $P_\eta = \text{diag}( 1, 1, 1, 1, -1, 1)$ as $\Sigma\to P_\eta \Sigma$, with which the global symmetry is enlarged to $O(6)/O(5)$~\cite{Frigerio:2012uc}. Under the parity $P_\eta$, $\eta$ is odd while all the SM particles are even. All the terms that are odd under $P_\eta$, including the Wess-Zumino-Witten terms~\cite{Wess:1971yu,Witten:1983tw,Chu:1996fr}, are forbidden. If $\eta$ is the lightest $P_\eta$-odd particle, it is stable and serves as a good DM candidate.

DM mass and the non-derivative Higgs-DM portal coupling are induced by any explicit shift-symmetry breaking effects. Since $\eta$ is a SM singlet, SM gauge bosons do not break DM shift symmetry. Though, explicit breaking effects can generally arise in fermion sectors, especially the top sector, as SM fermions are embedded in incomplete representations of $SO(6)$. For example, when the electroweak doublet $(t_L,b_L)^T$ and singlet $t_R$ are both embedded in fundamental representations of $SO(6)$, $(t_L,b_L)^T$ respects the DM shift symmetry while $t_R$ breaks it.
In order to fully restore DM shift symmetry in the top Yukawa sector, a vector-like softon field $X_{L,R}$ is introduced, where the right-handed component $X_R$ is embedded such that
the Lagrangian is invariant under the $SO(2)_d$ rotation $\mathcal{R}$ acting on the fifth and sixth components in accordance with the $\Sigma$ field as shown in Eq.~\ref{eq:sig}:
\begin{align}
\Sigma \to \mathcal{R} \Sigma,\ \  \Psi_{L,R}\to \mathcal{R} \Psi_{L,R},
\end{align}
and therefore $\Sigma^T \Psi_{L,R}$ is invariant, i.e.
\begin{equation}
\Sigma^T \Psi_{L,R}\to \Sigma^T \Psi_{L,R},
\end{equation}
where the fermionic fields are
\begin{align}
\Psi_L &= \frac{1}{\sqrt{2}}  ( i b_L, b_L, i t_L, -t_L, 0, 0 )^T,  \nonumber  \\
\Psi_R &= (0, 0, 0, 0, X_R, t_R )^T.
\end{align}
This rotation $\mathcal{R}$ defines the shift symmetry of $\eta$, i.e. $\eta$ is an exact Goldstone boson if the rotation remains an unbroken symmetry.
For any non-vanishing dependence of $\eta$ in the Yukawa sector, one can always perform an $SO(2)_d$ rotation to rotate it away.
Furthermore, the left-handed component $X_L$ is assumed as a $SO(6)$ singlet. One can introduce the mass term of $X_{L,R}$ as
\begin{equation}
\mathcal{L} \supset m_X \bar{X}_L X_R + \text{h.c.},
\end{equation}
which breaks $SO(2)_d$ explicitly and serves as the only source that gives nonzero DM mass and portal coupling $\lambda$, rendering $\eta$ a PNGB.

In the momentum space, the effective Lagrangian below the scale of strong dynamics is
\begin{align}
\mathcal{L}_{\text{eff}}=&\ \Pi_{L0} \bar{\Psi}_L p\!\!\!/ \Psi_L+\Pi_{R0}\bar{\Psi}_R p\!\!\!/ \Psi_R+ \bar{X}_L p\!\!\!/ X_L \nonumber\\
&+\Pi_{L1} \left( \bar{\Psi}_L \Sigma\right) p\!\!\!/ \left(\Sigma^T \Psi_L\right) + \Pi_{R1} \left( \bar{\Psi}_R \Sigma \right) p\!\!\!/ \left( \Sigma^T \Psi_R \right) \nonumber\\
&-\Pi_{t} \left( \bar{\Psi}_L \Sigma \right) \left( \Sigma^T \Psi_R \right) - m_X \bar{X}_L X_R + \text{h.c.}\ ,
\label{eff_Lag}
\end{align}
where $\Pi_{L0, L1}, \Pi_{R0,R1}, \Pi_t$ are the momentum-dependent form factors which are all calculable in concrete composite models. In the decoupling limit of $m_X\to\infty$, one can turn off the fermionic field $X$, this corresponds to the breaking of DM shift symmetry in conventional top sectors. Notice the terms $\left( \bar{\Psi}_L \Sigma\right) p\!\!\!/ X_L$ and $\left( \bar{\Psi}_R \Sigma\right) X_L$ are forbidden as they are odd under the dark parity.

\section{Effective Potential} \label{sec:potential}

In the following, we present a concrete composite model from which one can match to the effective Lagrangian in Eq.~\ref{eff_Lag} and calculate the scalar potential, where the leading terms are given by
\begin{align}
V(h,\eta) \simeq \frac{1}{2} m_h^2 h^2 + \frac{1}{2} m_\eta^2 \eta^2 + \frac{1}{4} \lambda_h h^4 +\frac{1}{4} \lambda_\eta \eta^4 + \frac{1}{2} \lambda h^2 \eta^2.
\label{eq:pot}
\end{align}
In particular, we demonstrate that the portal coupling $\lambda$ in the soft-breaking scenario is suppressed compared to its NDA value, which however can be recovered at the decoupling limit as $m_X\to\infty$.

We introduce composite resonances $Q$ and $S$ arising from strong dynamics, which are respectively in the fundamental representation and singlet of $SO(5)$, and construct the composite model following CCWZ~\cite{Coleman:1969sm,Callan:1969sn} with the Goldstone matrix $U=\exp{\left( i \frac{\sqrt{2}}{f} \pi^a T^a \right)}$, where $\pi^a$ are the Goldstone bosons and $T^a$ are the broken generators of $SO(6)/SO(5)$. Notice the Goldstone matrix realizes the $SO(6)$ symmetry nonlinearly.
Under the paradigm of partial compositeness~\cite{Kaplan:1991dc},
\begin{align}
\mathcal{L}_{\text{top}} &= i \bar{\Psi}_L D\!\!\!\!/\, \Psi_L + i \bar{\Psi}_R D\!\!\!\!/\, \Psi_R + i \bar{X}_L D\!\!\!\!/\, X_L  - m_X \bar{X}_L X_R   \nonumber \\
&+ \sum_{i=1}^{N_Q}{\bar{Q}_i \left( i D\!\!\!\!/\, + e\!\!\!/ - m_{Q_i} \right) Q_i} + \sum_{j=1}^{N_S}{ \bar{S}_j \left( i D\!\!\!\!/\, -m_{S_j} \right) S_j }            \nonumber\\
&+ \sum_{i=1}^{N_Q}{\left( \epsilon^i_{tQ} \bar{\Psi}^A_R U_{Aa} Q^a_{iL}+ \epsilon^i_{qQ} \bar{\Psi}^A_L U_{Aa} Q^a_{iR} \right)} \nonumber\\
&+ \sum_{j=1}^{N_S}{\left( \epsilon^j_{tS} \bar{\Psi}^A_R U_{A6} S_{jL} + \epsilon^j_{qS} \bar{\Psi}^A_L U_{A6} S_{jR} \right)}  +\text{h.c.}\ ,
\label{top-Lagrangian}
\end{align}
where $A=(1,\cdots,6), a=(1,\cdots,5)$, $N_{Q,S}$ indicate the layers of composite resonances, and various $\epsilon$'s are the mixing parameters between the elementary fields $\Psi_{L,R}$ and the composites $Q,S$. We consider the minimal scenario with only one layer of composites, i.e. $N_Q = N_S = 1$ in this work, which however is not viable in the models of hard shift symmetry breaking~\cite{Marzocca:2014msa}. The top mass is obtained after integrating out heavy composites and turning on the Higgs vacuum expectation value, e.g. in~\cite{Marzocca:2012zn, Panico:2015jxa}.

\begin{figure}[t]
\centering
\includegraphics[width=9cm]{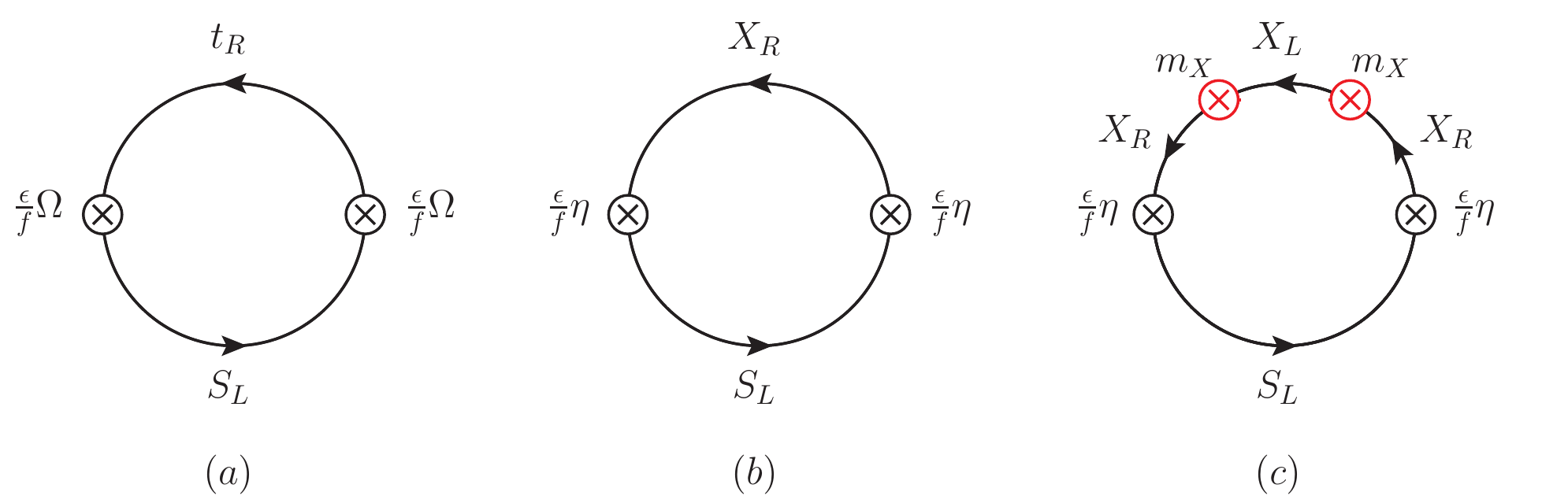}
\caption{Illustrative Feynman diagrams that give rises to the masses of $h,\eta$ via mixings with the composite resonance $S_L$, where $\Omega = \sqrt{f^2 - h^2 - \eta^2} $.}
\label{fig:loop-diagrams}
\end{figure}

As shown in Fig.~\ref{fig:loop-diagrams}, one consequence of imposing soft breaking is that the scalar $\eta$ does not suffer from the usual quadratic divergence. Because the dependence on $\eta$ vanishes in the quadratically divergent diagrams $(a)$ and $(b)$, which do not have $m_X$ insertions. Once $m_X$ is inserted, e.g. as in $(c)$, the diagrams with non-vanishing dependence of $\eta$ arise, and they are at most logarithmically divergent or finite. Notice we only consider the contribution coming from the composite $S$ in Fig.~\ref{fig:loop-diagrams}, likewise one can also include the contribution of the composite $Q$. Insertions of $m_S$ can further reduce the degree of divergence of the diagrams as shown in Fig.~\ref{fig:loop-diagrams}, however the terms of non-vanishing $\eta$ in the potential cannot arise without $m_X$ insertions.

\begin{figure}[t]
\centering
\includegraphics[width=8cm]{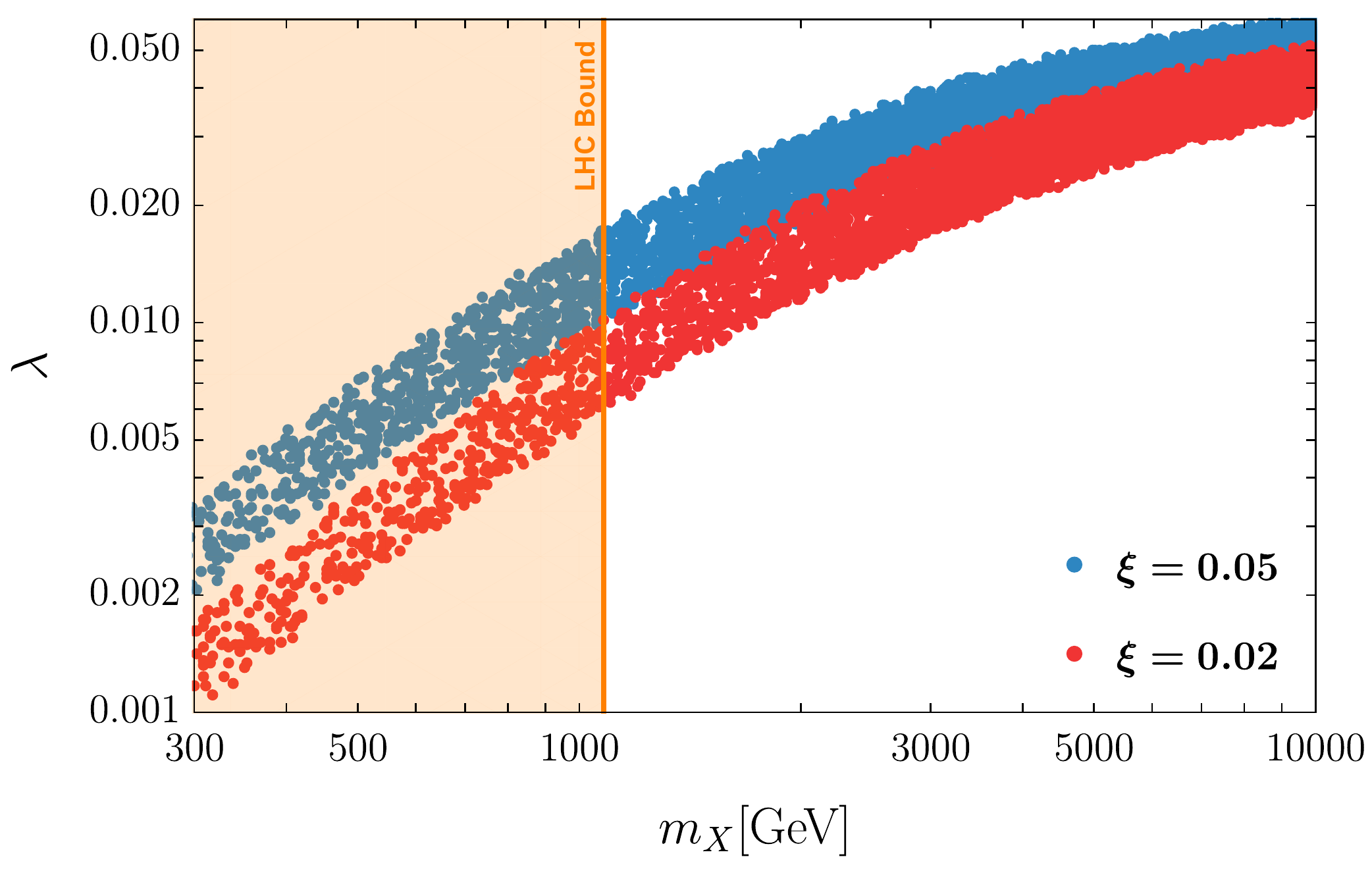}
\caption{Values of the portal coupling $\lambda$ as a function of the softon mass $m_X$, where the misalignment parameter $\xi= 0.02$ and $\xi = 0.05$, respectively. The shaded regime denotes the range of $m_X$ excluded by the LHC. The parameters in the model are scanned in the ranges of $m_{Q,S}\in\left[0,15f\right], m_X\in\left[0,10f\right], \epsilon \in \left[ 0, 2f \right], f_\rho\in\left[ f/\sqrt{2},2f\right]$, where $f_\rho$ comes from gauge sector~\cite{Marzocca:2014msa}.}
\label{fig:mX-lambda}
\end{figure}

In order to get rid of all the divergences and make the scalar potential fully calculable, one can impose Weinberg sum rules (WSRs)~\cite{Marzocca:2012zn,Pomarol:2012qf,Li:2019ghf}.
When $N_Q = N_S = 1$, one optimal choice can be
\begin{equation}
\epsilon_{qS} = -\epsilon_{qQ} = \sqrt{2}\epsilon_{tS} = \sqrt{2}\epsilon_{tQ} = \epsilon, 
\end{equation}
with which we calculate the form factors in Eq.~\ref{eff_Lag} and the scalar potential in Eq.~\ref{eq:pot}. Other choices of WSRs only modify the quantitative range of viable parameter space of various masses and mixings in Eq.~\ref{top-Lagrangian}, if all the low energy parameters in the scalar potential and the top mass are fixed. But the overall qualitative relations of various parameters will not change.
The portal coupling $\lambda$ is now calculable.
When the mixing $\epsilon$ is smaller than the symmetry breaking scale $f$, 
\begin{align}
\lambda \simeq & \frac{N_c \epsilon^4 }{4\pi^2 f^4}  \frac{ m_X^2 \left(m_Q+m_S\right)^2 }{ \left(m_X^2-m_Q^2\right) \left(m_X^2-m_S^2\right) \left(m_Q^2-m_S^2\right) }  \times    \nonumber \\
& \left( m_S^2\text{log}\frac{m_X^2}{m_Q^2} + m_Q^2\text{log}\frac{m_S^2}{m_X^2} + m_X^2\text{log}\frac{m_Q^2}{m_S^2} \right)\ ,
\end{align}
where $N_c=3$ is the QCD color factor from the top sector.
At the decoupling limit of $m_X \to \infty$, the conventional portal coupling at the NDA size~\cite{Marzocca:2014msa} is recovered, i.e.
\begin{equation}
\lambda_{\text{NDA}} \simeq \frac{N_c \epsilon^4 }{4\pi^2 f^4}  \frac{m_Q+m_S}{m_Q-m_S} \log \frac{m_Q^2}{m_S^2}\ .
\end{equation}
In contrast, when $m_X \lesssim m_{Q,S}$, an suppression factor for $\lambda$, defined as the ratio of the portal coupling in the shift-symmetry-soft-breaking scenario and that in the hard-breaking scenario, is obtained as
\begin{equation}
r \equiv \frac{\lambda}{ \lambda_{\text{NDA}} } \simeq \frac{m_X^2}{m_*^2} \log \frac{m_*^2}{m_X^2}\ , 
\label{eq:ratio}
\end{equation}
where $m_*$ denotes the overall mass scale of the strong dynamics, i.e. $m_Q \simeq m_S \simeq m_*$. The smaller the value of $m_X$, the smaller the portal coupling $\lambda$.
Without using the above approximation in the small $\epsilon$ limit, we scan the parameters in the composite model and plot the relation between $\lambda$ and $m_X$, as shown in
Fig.~\ref{fig:mX-lambda}. In the scan, we require the top mass, the Higgs mass, and the EW scale are reproduced.
The parameter $\xi\equiv v^2/f^2$ denotes the usual vacuum misalignment angle, where we choose $\xi=0.02$ and $\xi=0.05$ which are consistent with the bounds from precision measurements of various Higgs couplings~\cite{Li:2019ghf}. 
When the softon $X$ is heavy enough, the value of $\lambda$ approaches to the NDA value as $\lambda_{\text{NDA}}\simeq 0.065$. When $m_X$ is moderately heavier than the Large Hadron Collider (LHC) bound~\cite{Belyaev:2017vsx}, $\lambda$ can be much suppressed with respect to the NDA value. For example, $\lambda$ can be as small as $0.006$ when $\xi=0.02$. And the portal coupling will be more suppressed if the scale of strong dynamics (and the scale of $f$) is larger, as shown in Eq.~\ref{eq:ratio}.
Since the softon $X$ does not break the shift symmetry of the Higgs boson, the usual Higgs mass and quartic are reproduced.

\section{Dark Matter Phenomenology} \label{sec:DM-pheno}
\begin{figure}[t]
	\centering
	\includegraphics[width=8cm]{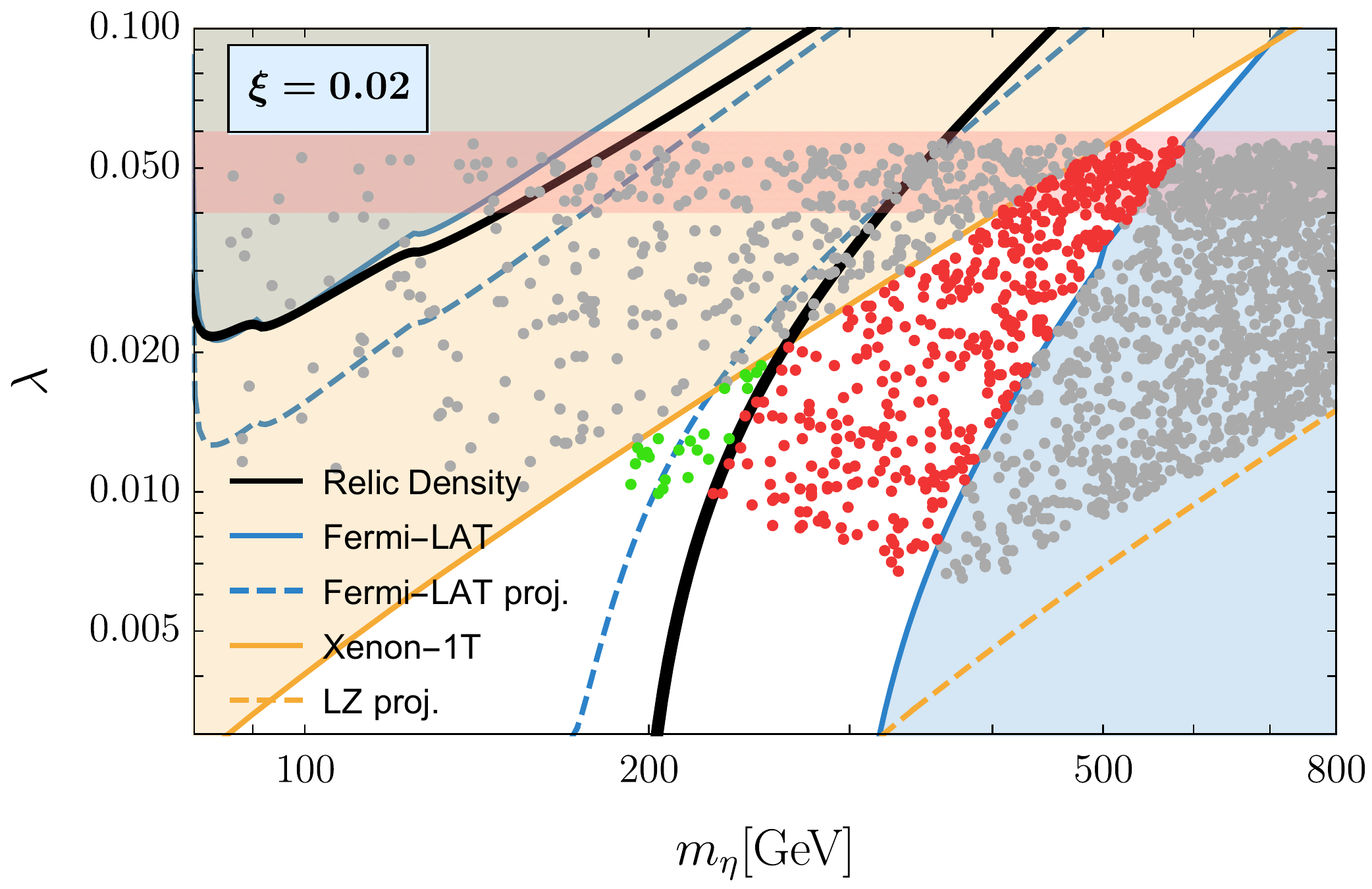}
	\caption{Constraints on the model parameters $(m_\eta, \lambda)$ plane when $\xi=0.02$, where only the unshaded red dots are fully viable under the current bounds from direct detection (yellow region), indirect detection (blue region), and relic abundance (black). All the other dots in the shaded region are excluded for various reasons, and the green dots indicate the region where DM are over produced. 
	The pink band denotes the NDA-size portal coupling given by~\cite{Marzocca:2014msa} with $N_S = 2N_Q = 2$. 
	}
	\label{fig:result}
\end{figure}

Since the mass of $\eta$ is at the EW scale, it is expected to exhibit traditional WIMP phenomenology, whereas it cannot be detected in current direct detection experiments. The numerical results are given in Fig.~\ref{fig:result}, where the details are summarized as follows.

In the early Universe, the relic abundance of $\eta$ is obtained by the freeze-out of the dark matter self-annihilation process, i.e. $\eta$ pair annihilating to SM particles.
As given by the Planck collaboration~\cite{Aghanim:2018eyx}, the relic abundance of $\eta$ is bounded as $\Omega_\eta h^2 \leq 0.120 \pm 0.001$. In the calculation of DM abundance, we assume the usual radiation dominance during DM freeze out. The derivative couplings, e.g. the operator $\mathcal{O}_1$ in Eq.~\ref{operators}, are dominant in the annihilation processes when $\eta$ is heavy. As a result, even with much suppressed portal coupling $\lambda$, the correct relic abundance can be reproduced~\cite{Frigerio:2012uc}. The softon field $X$ has only minor impact on determining the $\eta$ abundance. After the composite resonances are integrated out, the effective Lagrangian as in Eq.~\ref{eff_Lag} is obtained, which in particular contains the vertex $\eta \bar{t}_L X_R + \text{h.c.}$, where the softon $X$ is required to be heavier than $\eta$ such that $\eta$ is stable, and the searches for colored top partners also implies $m_X$ is above TeV scale~\cite{Belyaev:2017vsx}. Due to the heaviness of $X$, the annihilation channel $\eta\eta \to \bar{t} t$ induced by $X$ is suppressed, and the DM co-annihilation~\cite{Griest:1990kh} with $X$ can also be neglected. All the other processes changing the number density of $\eta$ are irrelevant.
Consequently, the freeze-out of $\eta$ in the early universe is dominated by the self-annihilaiton process in the Higgs portal. The relic abundance of $\eta$ reads,
\begin{equation}
	\Omega_\eta h^2 \simeq 0.12 \left( \frac{x_f}{24} \right) \left( \frac{3 \cdot 10^{-26} \text{cm}^3 \text{s}^{-1}}{\left< \sigma v_\text{rel} \right>_{x_f}} \right),
\end{equation}
where $x_f \equiv m_\eta/T_f$ and $T_f$ denotes the freeze-out temperature. $\left< \sigma v_\text{rel} \right>_{x_f}$ is the quantity of the thermally averaged cross section of DM annihilation at the freeze-out temperature. In this $SO(6)/SO(5)$ model, we have~\cite{Balkin:2017aep},
\begin{equation}
	\sigma v_\text{rel} \propto \left( \frac{s}{f^2} - 2\lambda \right)^2,
\end{equation}
where $s \approx 4m_\eta^2$ is the center-of-mass energy. The first term in the right-handed side of the equation is from operator $\mathcal{O}_1$ while the second from operator $\mathcal{O}_2$. The minus sign in the brackets indicates the cancellation of these two operators. Thus, for dark matter with certain mass, there could be two different values of $\lambda$ that correspond to correct relic abundance. This feature is shown explicitly in Fig.~\ref{fig:result}, where the observed DM abundance is denoted with the black bands.

As motivated in previous sections, $\lambda$ is much suppressed and therefore the bounds from current direct detection experiments are evaded. 
The spin-independent (SI) DM-nucleon cross section induced by operator $\mathcal{O}_2$ is given in Ref.~\cite{Cline:2013gha,Arcadi:2019lka}, which is the same as the usual Higgs portal singlet DM.
To be specific, it is given by,
\begin{equation}
	\sigma_\text{SI}^{\eta N} = \frac{1}{\pi} \left( \frac{m_N}{m_\eta+m_N} \right)^2 \left( \frac{Z f_p + (A-Z)f_n}{A} \right)^2.
\end{equation}
Here, we have $A=130$ and $Z=54$ for Xenon, and $m_N = (m_p+m_n)/2$ is the average nucleon mass. $f_{n,p}$ describe the coupling between DM and the necleons.
Substituting the values of these quantities, it reads,
\begin{equation}
	\sigma_\text{SI}^{\eta N} \simeq  5  \cdot 10^{-47} \text{cm}^2 \left( \frac{\lambda}{0.02} \right)^2 \left( \frac{300 \text{GeV}}{m_\eta} \right)^2 . 
\end{equation}
Such a value of DM-nucleon cross section lies well below the bounds of current direct detection experiments.
In addition, the softon $X$ only contributes to DM-nucleon scatterings at loop levels. When the tree-level DM-nucleon scatterings are suppressed due to the smallness of $\lambda$ as in our model, one-loop DM-nucleon scatterings can possibly dominate over tree-level ones. The yellow-shaded region is excluded by the current Xenon-1T experiment~\cite{Aprile:2018dbl}, and the reach of future LZ experiment~\cite{Akerib:2018lyp} is also shown by the yellow dashed line.
For completeness, we also show the constraints from indirect detection experiments, where the results from Fermi-LAT collaboration~\cite{Ackermann:2015zua,Fermi-LAT:2016uux} are used and we require the total thermal-averaged DM annihilation cross section to be smaller then the value reported by Fermi-LAT, assuming all the DM particles annihilate to the $\bar{b} b$ final state. In Fig.~\ref{fig:result}, the blue-shaded region is excluded by the current bounds~\cite{Ackermann:2015zua,Fermi-LAT:2016uux} and the future projection~\cite{Charles:2016pgz} is given by blue-dashed curve.
Numerically, we note that the relic abundance of $\eta$ and direct/indirect detection bounds are computed with the help of the public code \texttt{micrOMEGAs}~\cite{Belanger:2018mqt} and the model files are obatined with \texttt{FeynRules}~\cite{Alloul:2013bka}.
The constraints on DM mass and portal coupling $\lambda$ from invisible Higgs decay is not included here since we concentrate on the heacy DM region with $m_\eta > m_h/2$.

In addition, we present the result from parameter scan, as shown by the dots in Fig.~\ref{fig:result}.
We scanned randomly within the parameter range of $m_{Q,S}\in\left[0,15f\right], m_X\in\left[0,10f\right], \epsilon \in \left[ 0, 2f \right], f_\rho\in\left[ f/\sqrt{2},2f\right]$. $f_\rho$ is a parameter from the gauge sector~\cite{Marzocca:2014msa}.
Meanwhile, we require all the heavy resonances lie above the current LHC bounds, e.g.~\cite{Aaboud:2018pii,Sirunyan:2018yun,Belyaev:2017vsx,Liu:2018hum}, and the correct Higgs and top masses, as well as the electroweak scale are reproduced. The dots that do not meet these requisites are abandoned and not shown.
As is seen in Fig.~\ref{fig:result}, only the unshaded red dots are fully viable under the current bounds from direct detection (yellow region), indirect detection (blue region), and relic abundance (black). All the other dots in the shaded region are excluded for various reasons, and the green dots indicate the region where DM are over produced. In particular, compared to the NDA-size portal coupling in~\cite{Marzocca:2014msa} as shown by the pink band, the portal coupling in our model is much suppressed, and a large amount of viable parameter space, as represented by the unshaded red dots, is opened up, which can be tested in the near future.

\section{Concluding Remarks}

We propose soft-breaking mechanism for the shift symmetry of PNGB DM, which is still at the EW scale whereas the stringent bounds from current direct detection are evaded. We present a proof-of-concept model to demonstrate the usefulness of our mechanism. Generalizations to other models are straightforward. 
In addition, we notice that the similar soft-breaking mechanism can also be implemented to break the Higgs shift symmetry~\cite{Blasi:2019jqc}, where the top partner masses can be uplifted.
Implementing the soft breaking for both the Higgs and DM shift symmetries would lead to ``double" suppression for the non-derivative portal coupling, i.e. 
\begin{equation}
\frac{\lambda}{\lambda_{\text{NDA}}}\simeq  \frac{m_X^2}{m_*^2} \log \frac{m_*^2}{m_X^2} \cdot \frac{m_Y^2}{m_*^2} \log \frac{m_*^2}{m_Y^2}\ ,
\end{equation}
where $m_*$ is the typical scale of strong dynamics, $m_X$ and $m_Y$ are the softon masses for the breaking of the DM and Higgs shift symmetries, respectively. When the portal coupling is very suppressed, one-loop DM-nucleon scatterings will dominate over the tree-level ones in this class of models. We leave this for future investigation.

\section*{Acknowledgments}
L.X.X. would like to thank Roberto Contino, Rashmish K. Mishra, Alessandro Podo for valuable discussions on related topics, and Scuola Normale for warm hospitality when working on this project. This work is supported in part by the National Science Foundation of China under Grants No. 11635001, 11875072.

\onecolumngrid
\appendix
\section{From CCWZ Construction to Effective Potential}   \label{app:CCWZ_to_effective_potential}

Generators for $SO(6)/SO(5)$ group are chosen as follows,
\begin{align}
X_{i j}^{\hat{a}} &= -\frac{i}{\sqrt{2}} \left( \delta^{\hat{a} i} \delta^{6 j} - \delta^{\hat{a} j} \delta^{6 i} \right), \quad
T_{i j}^{\alpha} = -\frac{i}{\sqrt{2}} \left( \delta^{\alpha i} \delta^{5 j} - \delta^{\alpha j} \delta^{5 i} \right), \nonumber \\
T_{i j}^{a_{L, R}} &= -\frac{i}{2} \left( \frac{1}{2} \epsilon^{a b c} \left( \delta^{b i} \delta^{c j} - \delta^{b j} \delta^{c i} \right) \pm \left( \delta^{a i} \delta^{4 j} - \delta^{a j} \delta^{4 i} \right) \right), \nonumber
\end{align}
where $\hat{a}=1, \ldots, 5$ counts broken generators, while $a_{L, R}=1,2,3$ count generators of the subgroup $SU(2)_{L} \otimes SU(2)_{R} \cong SO(4) \subset SO(5)$ and $\alpha=1, \ldots, 4$ counts others. There are five Goldstones corresponding to five broken generators. Four of them living in fundamental representation of $SO(4)$ are identified as Higgs doublet and the other one which is $SO(4)$ singlet is considered as dark matter candidate. 
These Goldstone bosons can be parameterized by $U= \exp \left( i \sqrt{2} \frac{\pi^{\hat{a}}}{f} X^{\hat{a}} \right)$. We use $\pi^{\hat{a}}$ to represent the Goldstone bosons here. It is convenient to perform a field redefinition~\cite{Gripaios:2009pe} in the calculation: $\frac{ \sin (\pi / f) }{ \pi } \pi^{\hat{a}} \rightarrow \frac{\pi^{\hat{a}}}{f}$, where $ \pi \equiv \sqrt{ \pi^{\hat{a}} \pi^{\hat{a}} } $. To build Lagrangian, it is necessary to define CCWZ symbols $d_\mu$ and $e_\mu$,
\begin{equation}
d_{\mu}^{\hat{a}} X^{\hat{a}} + e_{\mu}^{a} T^{a} \equiv -i\left(U^{\dagger} D_{\mu} U\right). \nonumber
\end{equation}
$D_{\mu} $ is covariant derivative if a subgroup of $SO(6)$ is gauged. In our case, the gauged subgroup is the electroweak gauge group $SU(2)_L \otimes U(1)_Y$. We used $a = 1, \ldots, 10$ to count all unbroken generators of $SO(6)/SO(5)$. With the exception of the Wess-Zumino-Witten term~\cite{Wess:1971yu,Witten:1983tw,Chu:1996fr}, all terms can be constructed with $d_{\mu}$ and $ e_{\mu}$ symbols. In leading order, we have,
\begin{align}
\mathcal{L}_{\pi} &= \frac{f^{2}}{4} \operatorname{Tr}\left( d_{\mu}^{\hat{a}} d^{\mu \hat{a}}  \right).  \nonumber
\end{align}
Expanding this term, we can deduce the Lagrangian of Goldstones $h$ and $\eta$. 

For spin-1 resonances, we denote the resonances as $\rho_{\mu} \equiv \rho_{\mu}^{a} T^{a} \sim \bf{10}$ (adjoint representation) and $a_{\mu} \equiv a_{\mu}^{\hat{a}} X^{\hat{a} }\sim \bf{5}$ (fundamental representation). The Lagrangian is
\begin{align}
\mathcal{L}_{g} =  -\frac{1}{4} \operatorname{Tr} \left(\rho_{ \mu \nu } \rho^{\mu \nu} \right) + \frac{ f_{\rho}^2  }{2} \operatorname{Tr} \left( g_{\rho} \rho_{\mu} - e_{\mu} \right)^2
-\frac{1}{4} \operatorname{Tr} \left( a_{\mu \nu} a^{\mu \nu} \right) + \frac{  f_a^2  }{  2 \Delta^{2}  } \operatorname{Tr} \left( g_a a_{\mu} - \Delta d_{\mu} \right)^2. \nonumber
\end{align}
$f_{\rho,a}$ denote the decay constants for $\rho$ and $a$ respectively, and $g_{\rho,a}$ and $\Delta$ are constants. The field strengths and covariant derivatives are defined as
\begin{equation}
\rho_{\mu \nu}= \partial_{\mu} \rho_{\nu} - \partial_{\nu} \rho_{\mu} - i g_{\rho} \left[ \rho_{\mu}, \rho_{\nu} \right],
\quad
a_{\mu \nu}= \nabla_{\mu} a_{\nu}- \nabla_{\nu} a_{\mu}, 
\quad
\nabla_\mu=\partial_\mu-i e_\mu \nonumber
\end{equation}
Note that mass terms in $\mathcal{L}_{g}$ contain mixing terms between spin-1 resonances and electroweak gauge fields. Integrating these heavy resonances out, we will get effective Lagrangian for electroweak gauge bosons. In momentum space,
\begin{equation}
\mathcal{L}_g^{\text{eff}} = \frac{1}{2}  \left( g^{\mu \nu} - \frac{ p^\mu p^\nu}{p^2 } \right) \left( 2\Pi_{+-} W^+_\mu W^-_\nu + \Pi_{33} W^3_\mu W^3_\nu + \Pi_{BB} B_\mu B_\nu + 2\Pi_{3B} W^3_\mu B_\nu \right).
\end{equation}
where,
\begin{equation}
\Pi_{+-} = \Pi_{33} = \Pi_0 + \frac{h^2}{4f^2} \Pi^g_1,
\quad
\Pi_{BB} = \Pi_B + \frac{g^{\prime 2}}{g^2} \frac{h^2}{4f^2}\Pi^g_1,
\quad
\Pi_{3B} = -\frac{g^{\prime}}{g} \frac{h^2}{4f^2}\Pi^g_1. \nonumber
\end{equation}
$g,g^{\prime}$ are electroweak gauge coupling constants. In Euclidean space, the form factors are
\begin{equation}
\Pi_{0(B)} = Q^2 \left( 1+ \frac{ g^2(g^{\prime 2}) f_\rho^2 }{Q^2+m_\rho^2} \right),  
\quad
\Pi^g_1= g^2 \left( f^2 + 2 Q^2  \left(  \frac{f_a^2}{Q^2+m_a^2} - \frac{f_\rho^2}{  Q^2+m_\rho^2  } \right) \right). \nonumber
\end{equation}
Consequently, contribution to effective potential from the gauge sector reads
\begin{equation}
V_g(h) = \frac{3}{2} \int{ \frac{d^4Q}{(2\pi)^4} \text{log} \left[ \Pi^2_{+-}(\Pi_{33}\Pi_{BB} - \Pi_{3B}^2 ) \right]  }.
\label{effective_potential_gauge_appendix}
\end{equation}
$V_g(h)$ is $\eta$-independent since $\eta$ is singlet of electroweak gauge group. To render the integral finite, we need WSRs to eliminate divergence.
\begin{align}
\text{(WSR 1)}_{g}: \,  \frac{f^{2}}{2}+f_{a}^{2}-f_{\rho}^{2}=0, \quad
\text{(WSR 2)}_{g}: \,  f_{a}^{2} m_{a}^{2}=f_{\rho}^{2} m_{\rho}^{2}.
\label{WSR_gauge}
\end{align}
In parameter scan, we will use $f_{\rho}$ and $m_\rho$ as free parameters and deduce $f_a$ and $m_a$ accordingly with these two WSRs.

For fermion contribution, we add the expressions of the form factors in Eq.~\ref{eff_Lag} here,
\begin{gather}
\begin{aligned}
\Pi_{L0} &= 1 + \sum_{i=1}^{N_Q} \frac{\left|\epsilon^i_{qQ}\right|^2}{Q^2+m_{Q_i}^2}, 
&
\Pi_{L1} &= \sum_{j=1}^{N_S} \frac{\left|\epsilon^j_{qS}\right|^2}{Q^2+m_{S_j}^2} - \sum_{i=1}^{N_Q} \frac{ \left| \epsilon^i_{qQ} \right|^2}{Q^2+m_{Q_i}^2}, \\
\Pi_{R0} &= 1+ \sum_{i=1}^{N_Q} \frac{\left|\epsilon^i_{tQ}\right|^2}{Q^2+m_{Q_i}^2},
&
\Pi_{R1} &= \sum_{j=1}^{N_S} \frac{\left|\epsilon^j_{tS}\right|^2}{Q^2+m_{S_j}^2} - \sum_{i=1}^{N_Q} \frac{\left|\epsilon^i_{tQ}\right|^2}{Q^2+m_{Q_i}^2}, \nonumber \\
\end{aligned} \\
\Pi_{t} = \sum_{j=1}^{N_S} \frac{\epsilon^{*j}_{tS} \epsilon^j_{qS} m_{S_j}}{Q^2+ m^2_{S_j}} - \sum_{i=1}^{N_Q} \frac{\epsilon^{*i}_{tQ} \epsilon^i_{qQ} m_{Q_i}}{Q^2 + m_{Q_i}^2}.  
\label{form_factor}
\end{gather}
The scalar potential can be deduced from Eq.~\ref{eff_Lag} and the coefficients in Eq.~\ref{eq:pot} are integrals of the form factors.
\begin{align}
\label{eqt:coefficients_integral}
\mu^{(f)2}_h &= - \frac{N_c}{8 \pi^2 f^2} \int_0^\infty dQ^2 Q^2 \left( \frac{\Pi_{L1}}{\Pi_{L0}} + \frac{2\Pi_{R1}}{\Pi_{R0}} + \frac{\Pi_{t}^2}{Q^2 \Pi_{L0} \Pi_{R0}} \right), \nonumber \\
\lambda_h^{(f)} &= \frac{N_c}{4\pi^2 f^4} \int_{\mu^2}^\infty dQ^2 Q^2 \left( \frac{1}{4} \left( \frac{\Pi_{L1}}{\Pi_{L0}} + \frac{2\Pi_{R1}}{\Pi_{R0}} + \frac{\Pi_{t}^2}{Q^2 \Pi_{L0} \Pi_{R0}} \right)^2 + \frac{\Pi_{t}^2-Q^2\Pi_{L1} \Pi_{R1}}{Q^2 \Pi_{L0} \Pi_{R0}} \right),\nonumber \\
\mu_\eta^2 &= -\frac{N_c}{8 \pi^2 f^2} \int_0^\infty dQ^2 Q^2 \frac{m_X^2}{m_X^2+Q^2\Pi_X} \frac{\Pi_{R1}}{\Pi_{R0}}, \\
\lambda_\eta &= \frac{N_c}{16 \pi^2 f^4} \int_0^\infty dQ^2 Q^2 \left(  \frac{m_X^2 }{m_X^2+Q^2\Pi_X} \frac{\Pi_{R1}}{\Pi_{R0}} \right)^2, \nonumber \\
\lambda &= \frac{N_c}{16 \pi^2 f^4} \int_0^\infty dQ^2 Q^2 \frac{m_X^2}{m_X^2+Q^2\Pi_X} \left( \frac{\Pi_{R1}^{2}}{\Pi_{R0}^2}+ \frac{\Pi_{t}^{2}}{2 Q^{2} \Pi_{L0} \Pi_{R0}} \left(1+\frac{\Pi_{R1}}{\Pi_{R0}}\right) \right). \nonumber
\end{align}
We defined $\Pi_X=\Pi_{R0}+\Pi_{R1}$. To eliminate quadratic divergence in $\mu^{(f)2}_h$, we need to impose the following WSRs,
\begin{align}
\lim_{Q^2 \to \infty} Q^2 \frac{\Pi_{L1}}{\Pi_{L0}} = \sum_{i=1}^{N_Q} \left| \epsilon_{qQ}^i \right|^2 -\sum_{j=1}^{N_S} \left| \epsilon_{qS}^j \right|^2 = 0, \quad
\lim_{Q^2 \to \infty} Q^2 \frac{\Pi_{R1}}{\Pi_{R0}} = \sum_{i=1}^{N_Q} \left| \epsilon_{tQ}^i \right|^2 -\sum_{j=1}^{N_S} \left| \epsilon_{tS}^j \right|^2 = 0. \nonumber
\end{align}
The cancellation of logarithmic divergence in $\mu^{(f)2}_h$ requires,
\begin{equation}
\lim_{Q^2 \to \infty} Q^4 \left( \frac{\Pi_{L1}}{\Pi_{L0}} + \frac{2\Pi_{R1}}{\Pi_{R0}} \right) = \sum_{i=1}^{N_Q} \left( \left| \epsilon_{qQ}^i \right|^2 - 2\left| \epsilon_{tQ}^i \right|^2 \right) m_{Q_i}^2 -\sum_{j=1}^{N_S} \left( \left| \epsilon_{qS}^j \right|^2 - 2\left| \epsilon_{tS}^j \right|^2 \right) m_{S_j}^2 = 0. \nonumber
\end{equation}
Once these conditions are satisfied, all the UV divergences in the integrals Eq.~\ref{eqt:coefficients_integral} disappear.
In the case of $N_S=N_Q=1$ that is adopted in this work, one appropriate choice of WSRs that accomodates experimental constraints better is $$\epsilon_{qS} = -\epsilon_{qQ} = \sqrt{2}\epsilon_{tS} = \sqrt{2}\epsilon_{tQ} = \epsilon,$$ where $\epsilon$ is a constant.
Note that $\lambda_h^{(f)}$ is also IR divergent. It's because effective potential is singular at $h =0$. This issue can be cured by introducing another term $\Delta V = \delta h^{4} \log \left(h^{2} / f^{2}\right)$~\cite{Marzocca:2013fza}. It absorbs the singularity of Higgs quartic coupling constant. Nevertheless, we will simply cutoff this divergence with $\mu^2 = m_t^2$.

\twocolumngrid
\bibliography{SoftBreaking}

\end{document}